\documentclass[12pt]{article}
\usepackage{amsmath}
\usepackage{amsfonts}
\usepackage{amssymb}
\usepackage{hyperref}
\usepackage{eepic}
\newcommand{\Real}{\mathbb{R}}
\newcommand{\Kt}{\ensuremath{K_3}}
\newcommand{\VKt}{\ensuremath{V_{K_3}}}
\newcommand{\lh}{\ensuremath{l_{\rm H}}}
\newcommand{\lm}{\ensuremath{l_{\rm M}}}
\newcommand{\lii}{\ensuremath{l_{\rm II}}}
\newcommand{\li}{\ensuremath{l_{\rm I}}}
\newcommand{\lp}{\ensuremath{l_{\rm P}}}
\newcommand{\gh}{\ensuremath{g_{\rm H}}}
\newcommand{\gii}{\ensuremath{g_{\rm II}}}
\newcommand{\gi}{\ensuremath{g_{\rm I}}}
\newcommand{\gip}{\ensuremath{g_{\rm I'}}}
\newcommand{\gym}{\ensuremath{g_{\rm YM}}}
\newcommand{\giis}{\ensuremath{g_{\rm 6IIA}}}
\newcommand{\ghs}{\ensuremath{g_{\rm 6H}}}
\newcommand{\giib}{\ensuremath{g_{\rm IIB}}}
\newcommand{\giibs}{\ensuremath{g_{\rm 6IIB}}}
\def\be{\begin{equation}}
\def\ee{\end{equation}}
\def\ba{\begin{eqnarray}}
\def\ea{\end{eqnarray}}
\def\simlt{\stackrel{<}{{}_\sim}}

\makeatletter
\def\psheadings{\def\@oddfoot{}\def\@evenfoot{}
  \def\@oddhead{\hbox{}\hfill
        \makebox[.5\textwidth]{\raggedright\ignorespaces --\thepage{}--
        \hfill }}
  \def\@evenhead{\@oddhead}
  \def\subsectionmark##1{\markboth{##1}{}}
}
\renewcommand{\@makefnmark}{\mbox{$^{\ddagger\@thefnmark}$}}
\makeatother

\makeatletter
\newenvironment{scalepic}[3]
  {\begin{center} \stepcounter{equation} \def\@currentlabel{\theequation}      
   \label{#3}
   \begin{picture}(350,50)(-40,-25)%
   \put(-10,10){\makebox(0,0)[rb]{#1}}
   \put(-10,-10){\makebox(0,0)[rt]{#2}}
   \put(-10,0){\makebox(-50,0){$\updownarrow$}}
   \put(0,0){\vector(1,0){300}}
   \put(356,0){\makebox(0,0){(\theequation)}} }
  { \end{picture} \end{center} }
\newcommand{\scaleitem}[3]
 { \put(#1,0){ \put(0,-5){\line(0,1){10}}
               \put(0,10){\makebox(0,0)[b]{#2}}
               \put(0,-10){\makebox(0,0)[t]{#3}} }}
\newenvironment{doublescalepic}[4]
  {\begin{center} \stepcounter{equation}
   \def\@currentlabel{\theequation}      
   \label{#4}
   \begin{picture}(350,80)(-40,-55)%
   \put(-10,10){\makebox(0,0)[rb]{#1}}
   \put(-10,-10){\makebox(0,0)[rt]{#2}}
   \put(-10,0){\makebox(-50,0){$\updownarrow$}}
   \put(-10,-40){\makebox(0,0)[rt]{#3}}
   \put(-10,-30){\makebox(-50,0){$\updownarrow$}}
   \put(0,0){\vector(1,0){300}}
   \put(0,-30){\vector(1,0){300}}
   \put(356,-15){\makebox(0,0){(\theequation)}} }
  { \end{picture} \end{center} }
\newcommand{\downscaleitem}[3]
 { \put(#1,0){ \put(0,-35){\line(0,1){10}}
               \put(0,-20){\makebox(0,0)[b]{#2}}
               \put(0,-40){\makebox(0,0)[t]{#3}} }}
\newcommand{\doublescaleitem}[4]
 { \put(#1,0){ \put(0,-5){\line(0,1){10}}
               \put(0,10){\makebox(0,0)[b]{#2}}
               \put(0,-35){\line(0,1){10}}
               \put(0,-20){\makebox(0,0)[b]{#3}}
               \put(0,-40){\makebox(0,0)[t]{#4}} }}
\makeatother
\newcommand{\publititle}[8]
{ 
  \vspace*{-3cm}
  \begin{flushright} #1 \\ {\tt #2} \end{flushright}
  \vfill
  \begin{center}{\Large
    \bfseries #3}\end{center}
  \vskip 8mm
  \begin{center}{\large #4}\end{center}
  \begin{center}{\normalsize\sl #5}\end{center}
  \vskip 8mm     
  \nopagebreak 
  \noindent\begin{center} \parbox{.95\linewidth}{#6}\end{center}
  \vfill 
  \begin{flushleft} #7
  \end{flushleft}
  \hrule width 6.7cm \vskip.1mm
  {\small #8}
  \thispagestyle{empty}
  \clearpage
}

\begin{document}

\publititle{CPHT--S708-0299
}
{hep-th/9902055} 
{Low-Scale Closed Strings and their Duals$^\star$}
{I. Antoniadis and B. Pioline }
{Centre de Physique Th{\'e}orique$^\dagger$, 
Ecole Polytechnique,\\
{}F-91128 Palaiseau\\[+2mm]
{\small {\tt antoniad,pioline@cpht.polytechnique.fr}}
}
{\renewcommand{\baselinestretch}{1.2}\normalsize
We study large dimensions and low string scale in four-dimensional
compactifications of type II theories of closed strings at weak coupling.
We find that the fundamental string scale, 
together with all compact dimensions,
can be at the TeV, while the smallness of the string coupling
accounts for the weakness of gravitational interactions.
This is in contrast to the situation recently 
studied in type I theories, where the string scale
can be lowered only at the expense of introducing large transverse
dimensions felt by gravity only. As a result, in these type II strings, there
are no strong gravity effects at the TeV, and the main experimental signature
is the production of Kaluza-Klein excitations with gauge interactions.
In the context of type IIB theories, we find a new possibility providing a
first instance of large non-transverse dimensions at weak coupling:
two of the internal dimensions seen by gauge interactions
can be at the TeV, with the string scale and all other dimensions at
intermediate energies of the order of $10^{11}$ GeV, where gravity becomes
also strong. Finally, using duality, we provide a perturbative 
description for the generic case of large dimensions in the heterotic
string. In particular, we show that the two type II theories above 
describe the cases of one and two heterotic large dimensions. A new
M-theory derivation of heterotic-type II duality is instrumental
for this discussion.
}
{CPHT--S708-0299, February 1999, to appear in Nucl. Phys. B.}
{$^{\dagger}${\small Unit{\'e} mixte CNRS UMR 7644} \\
$^{\star}${\small Research supported in part by the EEC under the 
TMR contract ERBFMRX-CT96-0090.}
}

\clearpage
\psheadings
\renewcommand{\baselinestretch}{1.65}\normalsize

\section{Introduction}

Large dimensions are of particular interest in string theory
because of their possible use to explain outstanding physical
problems, such as the mechanism of supersymmetry breaking \cite{ia}, the
gauge hierarchy \cite{add,ab} and the unification of fundamental 
interactions \cite{w,ddg,cb}.

In the context of perturbative heterotic string theory, large TeV
dimensions are motivated by supersymmetry breaking which identifies
their size to the breaking scale. Although the full theory is
strongly coupled in ten dimensions, there are many quantities that
can be studied perturbatively, such as gauge couplings that are often
protected by non renormalisation theorems \cite{ia,ant}, as well as all soft
breaking terms because of the extreme softness of the supersymmetry
breaking mechanism, in close analogy with the situation at finite
temperature \cite{ia,soft}. 
An obvious question is whether there is some dual theory
that provides a perturbative description of the above models.

More recently, it was proposed that the observed gauge hierarchy
between the Planck and electroweak scales may be due to the existence
of extra large (transverse) dimensions, seen only by gravity
which becomes strong at the TeV \cite{add}. 
This scenario can be realised \cite{aadd,st} in the
context of a weakly coupled type I$^\prime$ string theory with a string
scale at the TeV \cite{l} and the Standard Model living on D-branes, 
transverse to $p$ large dimensions of size ranging from (sub)millimeter 
(for $p=2$) to a fermi (for $p=6$). As we will discuss below, these 
models are dual to the heterotic ones with 
$n=4$, 5 or 6 large dimensions. More precisely, the cases $n=4$ and $n=6$
correspond to $p=2$ and $p=6$, respectively, while the description of $n=5$
uses an anisotropic type I$^\prime$ theory with 5 large and one extra large
dimension.

In this work, we study large dimensions in the context of weakly
coupled type II string compactifications. One of the main
characteristics of these theories is that non-abelian gauge
symmetries appear non-perturbatively, even at very weak coupling, when
the compactification manifold is singular \cite{wi,ht2}. 
In particular, on the type IIA (IIB) side, they can be obtained from even (odd)
D-branes wrapped around even (odd) collapsing cycles. As we will show
in 
Section \ref{secii}, we find two novel possibilities which cannot be realised
perturbatively either in the heterotic or in type I constructions.

The first case consists of type IIA four-dimensional (4d)
compactifications with all internal radii of the order of the string
length, that can be as large as TeV$^{-1}$ due to a superweak string
coupling. Despite this, Standard Model gauge couplings remain of
order unity because their magnitude is determined by the geometry of
the internal manifold and not by the value of the string
coupling\footnote{This situation can also be realised in type I string theory
but only in six dimensions \cite{l,aadd}.}. This scenario offers an alternative
to the brane framework for solving the gauge hierarchy, with very different
experimental signals. There are no missing energy events from gravitons escaping
into the bulk \cite{add,aadd,grav},  while string excitations interact with
Planck scale suppressed couplings. As a result, the production of Kaluza-Klein
(KK) excitations with gauge interactions remain the only experimental probe
\cite{ia,abq}.  Furthermore, the problem of proton decay and flavor violation
becomes in principle much easier to solve than in low energy quantum gravity
models. This model will be shown in Section \ref{sechet} 
to be dual to the heterotic string with $n=2$ large dimensions.

The second case consists of type IIB theory with two large dimensions
at the TeV and the string tension at an intermediate scale of the
order of $10^{11}$ GeV. The string coupling is of order unity (but
perturbative), while the largeness of the 4d Planck mass is
attributed to the large TeV dimensions compared to the string length.
Gravity becomes strong at the intermediate scale and the main experimental
signal is again the production of KK gauge modes. However, the gauge
theory above the compactification scale is very different than in the
previous type IIA case. In fact,
this model offers the first instance of large radius along 
non-transverse directions and contains an energy domain where its
effective theory becomes six-dimensional below the string scale. 
This limit corresponds to a non-trivial infrared conformal point described by 
a tensionless self-dual string \cite{wi2}. 
This model is again dual to the heterotic string
with a single ($n=1$) large dimension at the TeV.

Of course, in the above two examples, one can increase the type IIA string
coupling or lower the type IIB string scale by introducing some extra large
dimensions transverse to the 5-brane where gauge interactions are localised.
In particular, we will show that the heterotic string with $n=3$ large TeV
dimensions is described by a type IIA compactification with a string scale and
two longitudinal dimensions at the TeV, four transverse dimensions at
the fermi scale, and order 1 string coupling.

The paper is organised as follows. In Section 2, we study TeV strings
and TeV dimensions in type II theories and we describe the first two examples
mentioned above. In Section 3, we review briefly the string dualities
among heterotic, type I and type II theories, and give the basic
relations we use in the sequel; in particular, we give a simple
(yet heuristic) derivation of heterotic--type IIA duality in the
framework of M-theory. In Section 4, we discuss large dimensions in
heterotic compactifications and provide a perturbative description of all
cases using heterotic -- type I or heterotic -- type II dualities. For
completeness, in Section 5, we examine large dimensions in type II
theories, and show how the heterotic theories of Section 4 are recovered.
Finally, Section 6 contains some concluding remarks.

\section{Type II theories with low string scale\label{secii}}

The Standard Model gauge interactions can be
in principle embedded within three types of four-dimensional
string theories, obtained by compactification of the ten-dimensional
heterotic, type II and type I theories. On the heterotic side, gauge
interactions appear in the perturbative spectrum and, like the gravitational
interactions, are controlled by the string coupling $\gh$. 
In type I theories, gauge interactions are described by open strings and 
confined on D-branes, whereas gravity propagates in the bulk; both interactions
are controlled again by the string coupling $\gi$, although gauge forces are
enhanced by a factor $1/\sqrt{\gi}$. In type II theories, the matching condition
forbids the existence of  non-abelian vector particles in the perturbative
spectrum; however, gauge interactions do arise non-perturbatively at
singularities of the $\Kt$-fibered Calabi-Yau manifold, from D2-branes (in type
IIA) or D3-branes (in type IIB) wrapping  the vanishing cycles\footnote{Here we
take as an example the well-understood case of $N=2$ supersymmetric
compactifications, since it already exhibits the main features of interest. Our
discussion carries over trivially to $N=1$ models obtained for instance by
freely acting orbifolds \cite{vw}.}. The gauge symmetry is dictated by the 
intersection matrix of the vanishing two-cycles in the $\Kt$ fiber,  whereas
extra matter arises from vanishing cycles localised at particular points on the
base \cite{mayr}. As a result, gauge interactions are localised on 5-branes at
the singularities with a gauge coupling given by a geometric modulus (the size
of the base in type IIA), whereas gravitational effects are still controlled by
the string coupling $\gii$.

More precisely, the gauge and gravitational kinetic terms in the effective
type IIA four-dimensional field theory are, in a self-explanatory notation:
\begin{equation}
S_{\rm IIA}=
\int d^4x\sqrt{-g}\left( \frac{1}{\giis^2}\frac{R_5R_6}{\lii^4}{\cal R}+
\frac{R_5R_6}{\lii^2}F^2\right)\, ,
\label{SII}
\end{equation}
where $\lii$ is the 
type II string length, $\giis$ is the six-dimensional string coupling, and
$R_5R_6$ is the two-dimensional volume of the base, along the 5-brane where
gauge fields are localised. For simplicity, we consider here the base to be a
product of two circles with radii $R_5$ and $R_6$. In eq. (\ref{SII}) and
henceforth we set all numerical factors to one, although we take them into
account in the numerical examples. Identifying the 
coefficient of $\cal R$ with the inverse Planck length $\lp^2$ and the
coefficient of $F^2$ with the inverse gauge coupling $\gym^2$, one gets:
\begin{equation}
\frac{1}{\gym^2}=\frac{R_5R_6}{\lii^2}\qquad,\qquad
\giis=\frac{1}{\gym}\frac{\lp}{\lii}\, .
\label{relii}
\end{equation}
Keeping the Yang-Mills coupling of order unity implies that $R_5R_6$ is of
order $\lii^2$, while $\giis$ is a free parameter that allows to move the
string length away from the Planck scale. As a result, one can take the type
II string scale to be at the TeV, keeping all compactification radii to be at
the same order of TeV$^{-1}$, by introducing a tiny string coupling of the
order of $10^{-14}$. We stress again that this situation cannot be realised in
heterotic or type I theories at weak coupling.

This scenario is very different from the TeV strings arising in type I theory,
where the string coupling is fixed by the 4d gauge coupling, while the
type I string scale is lowered by introducing large transverse dimensions
implying that gravity becomes strong at the TeV scale. Here, all internal
dimensions have string size (TeV$^{-1}$) and gravitational/string interactions
are extremely suppressed by the 4d Planck mass. As a result, Regge excitations
cannot be detected in particle accelerators and the main experimental signal
is the production of KK excitations of gauge particles, along the $(5,6)$
directions parallel to the 5-brane where gauge interactions are localised,
due to the singular character of the compactification manifold;
furthermore, these excitations come in multiplets of $N=4$ supersymetry, which
is recovered in the six-dimensional limit.
Quarks and leptons, on the other hand, do not in general
have KK excitations since matter fields are localised at
particular points of the base \cite{mayr}; they look similar to the twisted
fields in heterotic orbifold compactifications. Notice the similarity of these
predictions with those of heterotic string with large dimensions, despite its
strong 10d coupling \cite{ia}. The requirement of $N=4$ excitations was, there,
a way to keep the running of gauge couplings logarithmic above the
compactification scale. In fact, as we will show in Section
\ref{sechet}, the above type II models are dual to heterotic compactifications
with two large TeV dimensions. 

An obvious advantage of this scenario is that several potential
phenomenological problems, such as proton decay and flavor violations, appear
much less dangerous than in type I TeV strings,  since they are restricted
to the structure of KK gauge modes only, and one does not have to worry about
string excitations. Model building, on the other hand, becomes more involved
since it requires a deeper understanding of the gauge theory on the 5-brane
localised at the singular points of $\Kt$; due to the weakness of the string
coupling, the dynamics of the gauge theory is determined by classical string
theory in a strongly curved background, which can be analysed, for instance, in
the framework of geometric engineering \cite{mayr}.

A related question is the one of gauge hierarchy, which in the present context
consists of understanding why the type IIA string coupling is so small. The
technical aspect of this problem is whether string interactions decouple from
the effective gauge theory on the 5-brane, in the limit of vanishing coupling.
Although naively such a decoupling seems obvious, there may be subtleties
related to the non-perturbative origin of non-abelian gauge symmetries due to
the singular character of the compactification manifold. In fact, we would like
to argue that there are in general logarithmic singularities similar to the case
of having an effective transverse dimensionality $d_\perp=2$ in the
D-brane/type I scenario of TeV strings \cite{ab}.

The argument is based on threshold corrections to gauge and
gravitational couplings, that take the form:
\begin{equation}
\Delta = b\ln(\mu a) +\Delta_{\rm reg}(t_i)\, ,
\end{equation}
where the first term corresponds to an infrared (IR) divergent contribution,
regularised by an IR momentum cutoff $\mu$, $b$ is a numerical
$\beta$-function coefficient, and $\Delta_{\rm reg}$ is a function of
the moduli $t_i$, which in $N=2$ compactifications belong to vector
supermultiplets. For dimensional reasons, we have also introduced an
ultraviolet (UV) scale $a$, which naively should be identified with the type II
string length $\lii$. However because of the relation (\ref{relii}), in
supergravity units, $\Delta$ would acquire a dependence on the 4d string
coupling, which is impossible because it belongs to a neutral hypermultiplet
that cannot mix with vector multiplets in the low energy theory. This suggests
that $a$ should be identified with $\lp$ and thus, in string units,
there should be an additional contribution depending logarithmically on the
string coupling. This can also be understood either as a result of integration
over the massive charged states which have non perturbative origin, implying a
UV cutoff proportional to $(\lii\gii)^{-1}$, or from heterotic--type II duality
that we will discuss in Sections 3 and 4. Such a logarithmic dependence on the
string coupling  has also been observed in gravitational thresholds \cite{log}.

The logarithmic sensitivity on $\gii$ is very welcome
because it allows in principle the possible dynamical determination of the
hierarchy by minimising the effective potential. Note that the 
one-loop vacuum energy in non-supersymmetric type II vacua behaves
as $\Lambda\sim \lii^{-4}$ and thus is different from a quadratically
divergent contribution that would go as $(\lp l_{\rm str})^{-2}$. This should
be contrasted to the generic case of softly broken supersymmetry, as well as to
TeV type I strings with large transverse dimensions, where the
cancellation of this quadratic divergence implies a 
condition on the bulk energy density \cite{aadd}. 

Above, we discussed the simplest case of type II compactifications with string
scale at the TeV and all internal radii having the string size. In principle,
one can allow some of the $\Kt$ transverse directions to be large. From eq.
(\ref{relii}), it follows that the string coupling 
$\gii=\giis(\VKt/\lii^4)^{1/2}$, with $\VKt$ the volume of $\Kt$,
increases making gravity strong at distances 
$\lp(\VKt/\lii^4)^{1/2}$
larger than the Planck length. In particular, it can become strong at the TeV
when $\gii$ is of order unity. This corresponds to
$\VKt/\lii^4\sim 10^{26}$. It follows that in the isotropic case there are 4
transverse dimensions at a fermi, while in the anisotropic case 
$\VKt\sim R^\ell\lii^{4-\ell}$ the size of the transverse radii $R$ increases
with $\ell$ and reaches a micron for $\ell =2$. A more detailed analysis will
be given in Section 5.

We now turn on type IIB. As in type IIA, non-abelian gauge symmetries arise at
singularities of $\Kt$ from D3-branes wrapping around vanishing 2-cycles
times a 1-cycle of the base. Therefore, at the level of six dimensions,
they correspond to tensionless strings \cite{wi2}. 
The gauge and gravitational kinetic terms in the effective type IIB
four-dimensional action can be obtained from eq. (\ref{SII}) by T-duality with
respect, for instance, to the 6th direction:
\begin{equation}
R_6\to\frac{\lii^2}{R_6}\qquad\qquad 
\giis\to\giibs=\giis\frac{\lii}{R_6}\, .
\label{Tdual}
\end{equation}
One obtains
\begin{equation}
S_{\rm IIB}=
\int d^4x\sqrt{-g}\left( \frac{1}{\giibs^2}\frac{R_5R_6}{l_{II}^4}{\cal R}+
\frac{R_5}{R_6}F^2\right)\, ,
\label{SIIB}
\end{equation}
which leads to the identifications:
\begin{equation}
\frac{1}{\gym^2}=\frac{R_5}{R_6}\qquad,\qquad
\giibs=\frac{\lp}{\lii}\sqrt{\frac{R_5R_6}{\lii ^2}}=
\gym\frac{R \lp }{\lii^2}\, .
\label{reliib}
\end{equation}
Keeping the Yang-Mills coupling of order unity, now implies that $R_5$ and
$R_6$ are of the same order, $R_5=R_6/\gym^2\equiv R$, while $\giibs$ is a free
parameter. Obviously, in order to get a situation different from type IIA, $R$
should be larger than the string length $\lii$\footnote{Since $R$ corresponds
to a longitudinal direction, one can always choose $R>\lii$ by T-duality.}.
This corresponds to a type IIA string with large $R_5$ and small $R_6$ so that
$R_5R_6\sim\lii^2$. 

Imposing the condition of weak coupling 
$\gii=\giibs(\VKt/\lii ^4)^{1/2}\simlt 1$, we find
\begin{equation}
\lii\ge\sqrt{\gym R \lp}\, ,
\end{equation}
when all $\Kt$ radii have string scale size. As a result,
the type IIB string scale can be at intermediate energies
$10^{11}$ GeV when the compactification scale $R^{-1}$ is at the TeV.
The string coupling is then of order unity and gravity becomes strong at the
intermediate string scale. This brings back some worry on proton stability
although in a much milder form than in TeV scale quantum gravity models.
Of course, the string scale can be lowered by decreasing the string coupling
or by introducing large transverse dimensions in the $\Kt$ part. In the latter
case, gravitational interactions become strong at lower energies.

This result provides the first instance of a weakly coupled string theory with
large longitudinal dimensions seen by gauge interactions. In fact, as we will
show in Section 4, this theory describes heterotic compactifications with a
single large dimension. The existence of such dimension is motivated by
supersymmetry breaking in the process of compactification \cite{ia}. Note
that the physics above the compactification scale but below the type
IIB string scale is described by an effective six-dimensional theory of a
tensionless string \cite{wi2}. 
This theory possesses a non-trivial infrared dynamics at a
fixed point of the renormalisation group \cite{sei}\footnote{A similar
conclusion was obtained by studying the ultraviolet behavior of the
effective gauge theory \cite{fpt}.}. It would be interesting to study 
the dynamics of such theories in detail from the viewpoint of the reduced
four-dimensional gauge theory.

\section{Heterotic--type I--type II triality \label{trial}}

Here, we review briefly the basic ingredients of the dualities
between the heterotic, type I and type II string theories, 
that we will use in our subsequent analysis. The reader
with less string theoretical background may skip most of equations in this
part with the exception of the duality relations (\ref{heti}) and
(\ref{hetii}).

As is now well known, these three string theories are related by 
non-perturbative dualities, which take the simplest form in the
$N=4$ supersymmetric case of ($E_8\times E_8$ or $SO(32)$) 
heterotic theory compactified on
$T^6$, type II on $\Kt\times T^2$ and type I on $T^6$. We will
restrict our attention to this case, since it already exhibits the main
features we want to discuss. Heterotic--type I duality relates the
two ten-dimensional string theories with $SO(32)$ gauge group,
upon identifying \cite{pw}
\begin{equation}
\li=\gh^{1/2}\lh\ \,\qquad \gi=\frac{1}{\gh}\ ,
\label{heti}
\end{equation}
where $\li$ and $\lh$ denote the type I and heterotic string length.
The heterotic--type I duality itself holds in lower dimensions as well, 
and does not affect the physical shape or size of the 
compactification manifold. 

The $E_8\times E_8$ heterotic theory can in turn be perturbatively
related to the $SO(32)$ heterotic theory upon compactifying to
nine dimensions on a circle, since a $SO(1,17)$ boost transforms
the two even self-dual lattices into one another. More precisely,
the two theories are related by T-duality after breaking the
gauge symmetry to $SO(16)\times SO(16)$ on both sides by Wilson
lines \cite{gin}\footnote{This is in contrast to the unbroken $SO(32)$ 
or $E_8\times E_8$ phase, where each theory is self-dual under T-duality.}:
\begin{equation}
R_{\rm H'}=\frac{\lh^2}{R_{H}} \,\qquad 
g_{\rm H'} = g_{\rm H}\frac{\lh}{R_{H}}\ .
\label{hethet}
\end{equation}
where the prime refers to the $E_8\times E_8$ theory.
On the type I side, T-duality maps the theory with 32 D9-branes to 
a theory with 32 lower D$p$-branes, referred to in this work as type
I$^\prime_p$; the momentum states along the D9-branes are mapped to winding
states in the directions transverse to the D$p$-branes, and the action on 
the radii and coupling constant is the standard T-duality relation
$R\to\li ^2/R$,~$\gi\to \gi \li/R$.

On the other hand, heterotic--type IIA duality only arises in 6 dimensions or
lower,  and identifies again two theories with inverse six-dimensional
couplings $\ghs=\gh (\lh^4/V_4)^{1/2}$ and $\giis=\gii (\lii^4/\VKt)^{1/2}$,
with $V_4$ and $\VKt$ the volumes of $T^4$ and $\Kt$, respectively \cite{ht,wi}:
\begin{equation}
\lii=\giis \lh\ \,\qquad \giis=\frac{1}{\ghs}\ .
\label{hetii0}
\end{equation}
The identification of the scalar fields other than the dilaton 
can be obtained by decomposing the moduli space as 
\begin{equation}
\frac{SO(4,20)}{SO(4)\times SO(20)} \supset \left[ \Real^+ \times 
\frac{SO(3,19)}{SO(3)\times SO(19)} \right] \ .
\end{equation}
On the heterotic side, this involves choosing a preferred direction
of radius $R_1$ in $T^4$, parametrising the $\Real^+$ factor,
whereas on the type IIA side, the two factors occur naturally
as the overall volume and complex structure of $\Kt$, respectively. The volume 
of $\Kt$ in type IIA units is thus related to the radius $R_1$ 
in heterotic units as 
\begin{equation}
\label{rvk}
\left( \frac{R_1}{\lh} \right)^2 = \frac{\VKt}{\lii ^4}\ .
\end{equation}
Relating the other moduli fields requires a more precise
understanding of the geometry of $\Kt$ \cite{Aspinwall:1996mn}, but in
the following we will be able to obtain a partial identification
from the M-theory point of view.

Indeed, the above dualities can be understood in the M-theory
description\footnote{See for instance \cite{Obers:1998fb} for
a review.}, which subsumes the heterotic and type II descriptions
at strong coupling. The type IIA string theory then appears as M-theory
compactified on a vanishingly small circle of radius $R_s$ given by
\begin{equation}
R_s = \gii \lii\ ,\qquad \lm^3= \gii \lii ^3
\end{equation}
where $\lm$ denotes the eleven-dimensional Planck length \cite{wi,to};
the $E_8\times E_8$ heterotic string is obtained upon compactifying
on a segment $I$ of length $R_I$ given by analogous formulae:
\begin{equation}\
\label{hetm}
R_I = \gh \lh\ ,\qquad \lm^3= \gh \lh ^3\ .
\end{equation}
The two nine-branes at the end of the segment support the non-abelian
gauge-fields, whereas gravity propagates in the bulk \cite{hw}. In this
framework, a double T-duality symmetry $(R_1,R_2)$ $\to$ $(\lii ^2/R_2,\lii
^2/R_1)$ of type IIA acts as \cite{egkr,Obers:1998fb}
\begin{equation}
T_{ijk}~:~\tilde R_i=\frac{\lm^3}{R_j R_k}\ ,\ 
\tilde R_j=\frac{\lm^3}{R_i R_k}\ ,\ 
\tilde R_k=\frac{\lm^3}{R_i R_j}\ ,\ 
\tilde l_M^3=\frac{\lm^6}{R_i R_j R_k}\ ,\ 
\label{hmdf}
\end{equation}
where one of the radii $i,j,k$ corresponds to the
eleventh dimension; by eleven-dimensional general covariance, this
symmetry still holds for any choice of the three radii. 
As for the heterotic
T-duality $R_1\to \lh^2/R_1$, it translates into a symmetry
\begin{equation}
T_{Ii}~:~\tilde R_i=\frac{\lm^3}{R_I R_i}\ ,\ 
\tilde R_I=\frac{R_I ^{1/2} \lm^{3/2}}{R_i}\ ,\ 
\tilde l_M^3=\frac{\lm^{9/2}}{R_I ^{1/2} R_1}\ ,\ 
\label{hmdf2}
\end{equation}
where $R_i$ denotes the radius of any circular dimension and 
$R_I$ the length of the (single) segment direction.

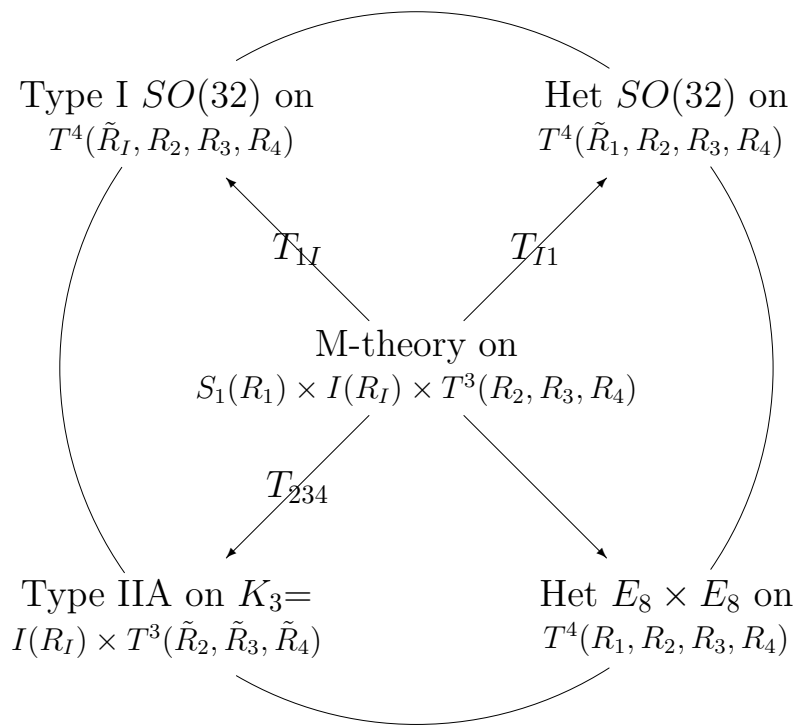
\begin{figure}
\begin{center}
\begin{picture}(300,300)(-150,-150)
\put(0,0){\arc{300}{-.6}{.6}}
\put(0,0){\arc{300}{2.53}{3.74}}
\put(0,0){\arc{300}{-2.1}{-1}}
\put(0,0){\arc{300}{1}{2.1}}
\put(20,20){\vector(1,1){60}}
\put(20,-20){\vector(1,-1){60}}
\put(-20,20){\vector(-1,1){60}}
\put(-20,-20){\vector(-1,-1){60}}
\put(0,9){\makebox(0,0){\large M-theory on}}
\put(0,-9){\makebox(0,0){$S_1(R_1)\times I(R_I)\times T^3(R_2,R_3,R_4)$}}
\put(105,114){\makebox(0,0){\large Het $SO(32)$ on}} 
\put(105,96){\makebox(0,0){$T^4(\tilde R_1,R_2,R_3,R_4)$ }}
\put(105,-96){\makebox(0,0){\large Het $E_8\times E_8$ on}} 
\put(105,-114){\makebox(0,0){$T^4(R_1,R_2,R_3,R_4)$}}
\put(-105,114){\makebox(0,0){\large Type I $SO(32)$ on}} 
\put(-105,96){\makebox(0,0){ $T^4(\tilde R_I,R_2,R_3,R_4)$}}
\put(-105,-96){\makebox(0,0){\large Type IIA on $K_3$=}} 
\put(-105,-114){\makebox(0,0){$I(R_I)\times T^3(\tilde R_2,\tilde R_3,\tilde R_4)$}}
\put(50,50){\makebox(0,0){\large $T_{I1}$}}
\put(-50,-50){\makebox(0,0){\large $T_{234}$}}
\put(-50,50){\makebox(0,0){\large $T_{1I}$}}
\end{picture}
\end{center}
\caption{Heterotic--Type I--Type II triality from M-theory\label{figtrial}}
\end{figure}

As depicted in Figure 1, we can now obtain the heterotic--type I--type II
relationships by interpreting the compactification of M-theory
on a manifold $S_1(R_1)\times I(R_I)\times T^3(R_2,R_3,R_4)$ in
various ways. (i) Considering $I(R_I)$ as the eleventh dimension
simply gives the $E_8\times E_8$ heterotic string on
$T^4(R_1,R_2,R_3,R_4)$ with string length and coupling given
by eq. (\ref{hetm}). (ii) Considering $S_1(R_1)$ as the eleventh
dimension gives type IIA on $I\times T^3$, or more properly 
type I$^\prime_8$ on $I\times T^3$; (iii) By T-duality $T_{I1}$ along the 
segment $I$, the theory (i) translates into heterotic $SO(32)$,
whereas the theory (ii) turns into type I: it
is straightforward to check that these two are related by the
duality relations (\ref{heti}) \cite{hw}. (iv) if we perform a $T_{234}$
duality on the torus $T^3$ before identifying $S_1(R_1)$ with the
eleventh dimension, we obtain a type IIA theory compactified on
a four-dimensional manifold 
$I(R_I) \times T^3(\tilde R_2,\tilde R_3,\tilde R_4)$, which
by (i) is the same as the $E_8\times E_8$ heterotic string on 
$T^4(R_1,R_2,R_3,R_4)$. It is easy to check that these two
theories are related by the heterotic--type II duality relations
(\ref{hetii0},\ref{rvk}). It is therefore tempting to identify
\begin{equation}
\Kt = I(R_I) \times T^3(\tilde R_2,\tilde R_3,\tilde R_4) \ ,
\label{K3}
\end{equation}
where the type II parameters are related to the heterotic ones as
\begin{subequations}
\label{hetii}
\begin{eqnarray}
\lii &=& \frac{\gh \lh ^3}{\sqrt{R_1 R_2 R_3 R_4}}\ ,\qquad
\gii= \frac{\sqrt{R_1 ^3 R_2 R_3 R_4}}{\gh \lh ^3}\ ,\qquad\label{hetii1}\\
R_I&=&\gh \lh\ ,\qquad 
\tilde R_i=\frac{\gh \lh ^3}{R_j R_k}\ ,\ i,j,k=2,3,4\ .\label{hetii2}
\end{eqnarray}
\end{subequations}
In units of the respective string length, this is\footnote{
This mapping was independently obtained by Polchinski as referred to
in \cite{dine}.}
\begin{subequations}
\begin{eqnarray}
\left(\frac{R_I}{\lii}\right)^2 &=& \frac{R_1 R_2 R_3 R_4}{\lh^4}\ ,\\
\left(\frac{\tilde R_2}{\lii}\right)^2 &=& \frac{R_1 R_2}{R_3 R_4}\ ,\quad
\left(\frac{\tilde R_3}{\lii}\right)^2 = \frac{R_1 R_3}{R_2 R_4}\ ,\quad
\left(\frac{\tilde R_4}{\lii}\right)^2 = \frac{R_1 R_4}{R_2 R_3}\ ,\quad
\end{eqnarray}
\end{subequations}
where we recognize a triality transformation in the $[SO(4)\times
SO(4)] \backslash SO(4,4)$ subspace of the moduli space.
Even though (\ref{K3}) is not a proper $\Kt$ surface
(for one thing it is not simply connected), it still is na bona fide
compactification manifold, albeit singular. Indeed, it
has been argued that such a ``squashed'' shape arises in the 
decompactification limit of the heterotic torus $T^4$  
at the $E_8 \times E_8$ enhanced
symmetry point \cite{Aspinwall:1998eh}. This representation of
$\Kt$ will turn out to be very convenient in the discussion of large 
radius behaviour of heterotic and type II theories in the sequel.

\section{Large dimensions in heterotic string \label{sechet}}

Here we consider the heterotic string compactified in four dimensions with 
a certain number $n$ of large internal dimensions. Keeping the
four-dimensional gauge coupling $\gym$ of order unity,
the heterotic theory is strongly coupled with a ten-dimensional string coupling
and four-dimensional Planck length
\begin{equation}
\gh=\gym \left(\frac{R}{\lh}\right)^{n/2}\gg 1\ ,\qquad
\lp=\gym \lh 
\end{equation}
where $R$ is the common radius of the large dimensions, while the remaining
$6-n$ are assumed to be of the order of the string length $\lh$. For $n<6$, the
distinction between the $SO(32)$ and $E_8\times E_8$ heterotic
theories is irrelevant, since a T-duality (\ref{hethet}) along a
heterotic-size direction converts one into another; we will
therefore omit this distinction until we discuss the $n=6$ case,
where such a dualisation is no longer innocuous.

In order to obtain a perturbative description of this theory, we consider
first its type I dual obtained through the relations (\ref{heti}).
The physical radii of the internal manifold are
unaffected by this duality. In particular, there are still $6-n$ dimensions
of size $\lh$, and $n$ of size $R$. We therefore have the following 
type I string length and coupling
\begin{equation}
\li=\gym^{1/2} R^\frac{n}{4} \lh^{1-\frac{n}{4}}\ ,\qquad
\gi=\frac{1}{\gym}\left(\frac{\lh}{ R}\right)^\frac{n}{2}\ ,
\end{equation}
or more explicitely
\begin{equation}
\mbox{Type I:}\quad
\left\{
\begin{array}{lcccccc}
n=1,2,3,4: & \lh & < & \li & < &R \\
n=5,6: & \lh & < & R   & < &\li
\end{array} \right.
\end{equation}
In both cases, there are dimensions ($6-n$ or 6 respectively) with size
smaller than the type I string length, which should be T-dualised in order to
trade light winding modes for Kaluza-Klein (KK) field-theory states. In so
doing, we move to a type I$^\prime$ description where the gauge interactions are
localised on D-branes (extended in $3+n$ or 3 spatial directions,
respectively). Using the standard ($\hat R= \li^2/R$,
$\hat\gi=\gi \li/R$) T-duality relations, we obtain
the dual radii
\begin{equation}
\hat \lh=\gym R^\frac{n}{2} \lh^\frac{1-n}{2}\ ,\quad
\hat R=\gym R^\frac{n-2}{2} \lh^\frac{2-n}{2}
\end{equation}
and the hierarchies
\begin{equation}
\mbox{Type I$^\prime$:}\quad
\left\{
\begin{array}{lcccccc}
n=1,2~:~ \gip=\gym^\frac{4-n}{2}
\left(\frac{R}{\lh}\right)^\frac{n(4-n)}{4}\ ,& \li & < & \hat \lh & < &R \\
n=3,4~:~ \gip=\gym^\frac{4-n}{2}
\left(\frac{R}{\lh}\right)^\frac{n(4-n)}{4}\ ,& \li & < & R & < &\hat \lh \\
n=5,6~:~ \gip=\gym^2\ , & \li & < & \hat R   & < &\hat \lh
\end{array} \right.
\label{ip}
\end{equation}
where the T-dualised hatted radii correspond to transverse dimensions.

In the cases $n=1,2,3$, type I$^\prime$ theory is also strongly coupled, as
seen from eqs. (\ref{ip}), which is a consequence of the fact that the internal
longitudinal directions of the D-branes (of size $R$) are larger than the type
I$^\prime$ string length. In the cases $n=5,6$ however, the type I$^\prime$
theory does offer a perturbative description of the theory of interest, where
the gauge interactions are confined on D3-branes with large transverse
dimensions of size $\hat R$ ($n$ of them) and $\hat \lh$ ($6-n$ of them).
This is depicted in the following diagrams:
\begin{scalepic}{Het $n=5$}{I$^\prime_3, \gip=\gym^2$}{scal0}
\scaleitem{20}{1}{$\lh,R_6$}
\scaleitem{120}{}{R}
\scaleitem{145}{$\gym^{1/2} R^{5/4}$}{$\li$}
\scaleitem{170}{~~~~\qquad$\gym R^{3/2}$}{~~~~\qquad$\hat R_{1,2,3,4,5}$}
\scaleitem{270}{$\gym R^{5/2}$}{$\hat R_6$}
\end{scalepic}
\begin{scalepic}{Het SO(32) $n=6$}{I$^\prime_3, \gip=\gym^2$}{scal00}
\scaleitem{20}{1}{$\lh$}
\scaleitem{120}{R}{}
\scaleitem{170}{$\gym^{1/2} R^{3/2}$}{$\li$}
\scaleitem{220}{$\gym R^2$}{$\hat R_{1,2,3,4,5,6}$}
\end{scalepic}
In particular, the $SO(32)$ 
heterotic string with $n=6$ large dimensions, say at $10^8$
GeV, is dual to a type I$^\prime$ with string tension at the TeV and six
transverse dimensions at 0.1 fermi. This is one of the examples that were
treated recently in the context of TeV strings \cite{aadd}. 
The type I threshold appears in the strongly coupled heterotic theory below 
the KK scale of $10^8$ GeV \cite{ckm}, which is then identified with 
the scale of the
(superheavy) type I$^\prime$ winding states around the fermi-size 
transverse dimensions.

In the $n=4$ case, two distinct type I$^\prime$ perturbative descriptions
are possible, due to the proximity of the radius $R$ of the four large
dimensions with the type I string scale $\li=\gym^{1/2}R$. For $\gym<1$, it is
sufficient to T-dualise the two  directions of size $\lh$, resulting in a
D7-brane type I$^\prime$ description with string coupling unity, which can
be lowered by increasing for instance the size of the two small dimensions
slightly above the heterotic length. For $\gym>1$ on the other hand,
one should T-dualise also the remaining four directions, resulting in a D3-brane
type I$^\prime$ description as above. In both cases, the type I$^\prime$ scale
is close to the size of the four heterotic large dimensions, say at the TeV
scale. This provides another example of type I$^\prime$ TeV 
strings \cite{aadd}.
The gauge interactions are confined on D-branes transverse to two large
dimensions of (sub)millimeter-size. The type I threshold now appears at the
same order
as the KK scale (at the TeV) \cite{aq}, while the heterotic scale -- which is
also the KK scale of the remaining two dimensions -- is identified
with the mass
of the type I$^\prime$ winding modes around the two millimeter-size dimensions.
This model is of particular interest, because it 
offers a possibility to keep the
apparent unification of gauge couplings close to the heterotic scale, 
due to the
logarithmic sensitivity of the gauge theory on the brane with respect to the
size of the two-dimensional transverse space
\cite{cb,ab}.
\begin{scalepic}{Het $n=4$}{I$^\prime_7, \gip=1$}{scal01}
\scaleitem{20}{1}{$\lh$}
\scaleitem{100}{$\gym^{1/2} R$}{$\li$}
\scaleitem{140}{$R$}{$R_{1,2,3,4}$}
\scaleitem{260}{$\gym R^{2}$}{$\hat R_{5,6}$}
\end{scalepic}
\begin{scalepic}{Het $n=4$}{I$^\prime_3, \gip=\gym^2>1$}{scal02}
\scaleitem{20}{1}{$\lh$}
\scaleitem{140}{$R$}{}
\scaleitem{170}{$\gym^{1/2} R$}{$\li$}
\scaleitem{210}{$\gym R$}{$\hat R_{1,2,3,4}$}
\scaleitem{260}{$\gym R^{2}$}{$\hat R_{5,6}$}
\end{scalepic}

In order to obtain a perturbative description for the cases $n=1,2,3$, we now
consider the type IIA dual of the original heterotic theory. As described in 
Section \ref{trial}, the heterotic--type IIA duality selects four preferred 
dimensions of radii $R_{1,2,3,4}$ on the 
heterotic side, while at the $E_8\times E_8$ enhanced symmetry point 
the compactification manifold for the type IIA string takes the simplified
 form
\begin{equation}
\Kt\times T^2 = \left[ I(R_I) \times T^3(\tilde R_2,
\tilde R_3,\tilde R_4) \right] 
\times T^2( R_5,R_6)\ .
\label{ktt}
\end{equation}
The remaining two-torus of radii $R_5,R_6$ is common to both descriptions,
which also have the same four-dimensional gauge coupling and Planck mass
\begin{subequations}
\begin{eqnarray}
\frac{1}{\gym^2}&=&\frac{R_1 R_2 R_3 R_4 R_5 R_6}
{\gh^2 \lh ^6}=\frac{R_5 R_6}{\lii ^2}\ ,\\
\frac{1}{\lp^2}&=&\frac{R_1 R_2 R_3 R_4 R_5 R_6}{\gh^2 \lh ^8}=
\frac{R_I \tilde R_2 \tilde R_3 \tilde R_4 R_5 R_6}{\gii^2 \lii ^8}\ .
\end{eqnarray}
\end{subequations}
In terms of these quantities, the duality map (\ref{hetii}) takes the form
\begin{subequations}
\begin{eqnarray}
\lii&=&\gym \sqrt{R_5 R_6}\ ,\qquad
\gii=\frac{1}{\gym} \frac{R_1}{\sqrt{R_5 R_6}} \\
R_I&=&\gh \lh\ ,\qquad 
\tilde R_i=\frac{\gh \lh ^3}{R_j R_k}\ ,\ i,j,k=2,3,4
\end{eqnarray}
\end{subequations}
As we mentioned in the previous Sections, 
the four $\Kt$ directions corresponding
to $R_I$ and $\tilde R_i$ are 
transverse to the 5-brane where gauge interactions are
localised.

In order to obtain a weakly coupled type II description, we therefore 
need to carefully arrange the choice of the $n$ large dimensions on the 
heterotic side. For instance, in the $n=1$ case, choosing 
$R_1$ as the large radius results in a strongly coupled type II theory
with $\gii\sim R$ in units of the heterotic string length; 
choosing $R_2$ (or $R_3,R_4$)
as the large dimensions gives a type II dual with moderate coupling
$\gii\sim 1/\gym$, but with radii $\tilde R_3,\tilde R_4 \sim 1/\sqrt{R}$
much smaller than the string length; after T-dualisation along these 
directions, the theory becomes strongly coupled. The last option is
to take $R_5$ (or $R_6)$ as the large radius, which yields a weakly
coupled type II dual string with $\gii\sim 1/\gym \sqrt{R}$ and
$\lii \sim \gym \sqrt{R}$; after T-dualizing the heterotic-size direction 
$R_6$, we obtain a weakly coupled type IIB description with hierarchy
\begin{scalepic}{Het $n=1$}{IIB $\gii=1$}{scal1}
\scaleitem{20}{1}{$\lh$}
\scaleitem{120}{$\gym R^{1/2}$}{$\lii,R_{I,\tilde 2,\tilde 3,\tilde 4}$}
\scaleitem{220}{$\gym^2 R$}{$\hat R_6$}
\scaleitem{250}{$R$}{$R_5$}
\end{scalepic}
where we denoted by $\hat R_6$ the radius of the T-dual sixth
dimension. This is one of the models discussed in Section \ref{secii}, with
two radii at the TeV and a string scale at intermediate energies $10^{11}$ GeV.

In the $n=2$ case, the same reasoning leads to choosing the radii
$R_5,R_6\sim R$ as the large heterotic dimensions, and gives a weakly coupled
type IIA description
\begin{scalepic}{Het $n=2$}{IIA $\gii=\frac{1}{\gym R}$}{scal2}
\scaleitem{20}{1}{$\lh$}
\scaleitem{170}{$\gym R$}{$\lii,R_{I,\tilde 2,\tilde 3,\tilde 4}$}
\scaleitem{220}{$R$}{$R_{5,6}$}
\end{scalepic}
without need of T-dualizing any direction. This is the other type II model
discussed in Section \ref{secii}, with string scale and all internal dimensions
at the TeV, and with an infinitesimal string coupling $10^{-14}$ accounting for
the largeness of the four-dimensional Planck mass.

In the cases $n=3,4,5,6$, we choose the directions of radii
$R_1,R_5,R_6\sim R$ as three of the large heterotic dimensions,
and for $n>3$ also switch on $n-3$ large dimensions
in the heterotic $T^3(R_2,R_3,R_4)$ torus. The type II
dual has string length $\lii=\gym R$ and coupling $\gii=1/\gym$,
while the $\Kt$ manifold has size (in heterotic units):
\vspace*{3mm}
\begin{equation*}
\begin{array}{|c|c|c|c|c|c|}
\hline
n & R_I & \tilde R_2 &\tilde R_3 &\tilde R_4 \\ \hline
3 & \gym R^{3/2} &\gym R^{3/2} &\gym R^{3/2} &\gym R^{3/2} \\
4 & \gym R^2 &\gym R^2 &\gym R &\gym R \\
5 & \gym R^{5/2} &\gym R^{3/2} &\gym R^{3/2} &\gym R^{1/2} \\
6 & \gym R^3 &\gym R &\gym R &\gym R\\
\hline
\end{array}
\end{equation*}

\noindent Except for the $n=5$, where the existence of the small radius
$\tilde R_4$ implies strong coupling after T-duality, all these cases correspond
to a  weakly coupled type II dual. In the $n=3$ case, the type II dual 
provides a perturbative description of the heterotic theory 
that could not be reached on the type I side:
\begin{scalepic}{Het $n=3$}{IIA $\gii=\frac{1}{\gym}$}{scal3}
\scaleitem{20}{1}{$\lh$}
\scaleitem{120}{$\gym R$}{$\lii$}
\scaleitem{150}{$R$}{$R_{5,6}$}
\scaleitem{220}{$\gym R^{3/2}$}{$R_{I,\tilde 2,\tilde 3,\tilde 4}$}
\end{scalepic}
This is the type II model discussed in Section \ref{secii} with string scale
and two longitudinal dimensions at the TeV, and an isotropic $\Kt$ with 4
transverse directions at a fermi.

In the $n=4$ case, the type II dual theory provides a perturbative
description, alternative to the type I$^\prime$. 
The type II dual string has
the same scale hierarchy as the type I, up to factors of $\gym$:
\begin{doublescalepic}{I$^\prime_7, \gip=1$}{Het $n=4$}
{IIA $\gii=\frac{1}{\gym}$}{scal4}
\doublescaleitem{20}{$\lh$}{1}{$\lh$}
\downscaleitem{110}{$\gym R$}{$\lii,R_{\tilde 3,\tilde 4}$}
\scaleitem{150}{$\li$}{$\gym^{1/2} R$}
\doublescaleitem{180}{$R_{1,2,3,4}$}{$R$}{$R_{5,6}$}
\doublescaleitem{250}{$\hat R_{5,6}$}{$\gym R^{2}$}{$R_{I,\tilde 2}$}
\end{doublescalepic}
These two models should provide equivalent perturbative descriptions of the
same theory.

In the $n=6$ case, we now obtain a weakly coupled description
of the $E_8\times E_8$ heterotic string with $n=6$ large radii
as a type IIA string with string length $\lii=\gym R$:
\begin{scalepic}{Het $E_8\times E_8, n=6$}{IIA $\gii=\frac{1}{\gym}$}{scal5}
\scaleitem{20}{1}{$\lh$}
\scaleitem{80}{$\gym R$}{$\lii,R_{\tilde 2,\tilde 3,\tilde 4}$}
\scaleitem{120}{$R$}{$R_{5,6}$}
\scaleitem{270}{$\gym R^3$}{$R_I$}
\end{scalepic}
Due to the occurrence of gravitational KK states\footnote{Note that these
excitations are not stable due to the lack of momentum conservation along the
interval $I(R_I)$.} at the scale
$R_I$,  the type II string tension as well as the heterotic compactification
scale cannot be lower than $10^8$ GeV, corresponding to the bound $R_I\simlt 1$
mm. This situation should be contrasted with the case of the $SO(32)$ heterotic 
string with $n=6$ large radii, which admits a perturbative
description (\ref{scal00}) as a type I string with length
$\li=\gym^{1/2} R^{3/2}$. The bound on $R$ still
applies, corresponding now to a type I string scale at a TeV, 
and six transverse dimensions at 0.1 fermi.
Note that the difference between the type I and type II string scales
does not lead to any inconsistency, since the two
perturbative descriptions are not simultaneously possible.

\section{Large dimensions in type II theories and their duals\label{seciid}}
Having discussed the large radius behaviour of the dual heterotic
theory, we now reconsider the type IIA models we introduced in
Section \ref{secii}, and discuss their dual descriptions.
We therefore consider type IIA theory, compactified on the simplified
model (\ref{ktt}) of $\Kt\times T^2$, with a weak string coupling
$\gii\ll 1$, two string-size directions $R_5,R_6\sim \lii$ and
possibly $\ell$ large transverse directions of size $R\gg \lii$ within $\Kt$.
This theory is then identified to a strongly coupled heterotic string 
compactified on $T^6$, with parameters simply obtained by inverting
eq. (\ref{hetii}): 
\begin{subequations}
\label{iihet}
\begin{eqnarray}
\lh &=& \frac{\gii \lii ^3}{\sqrt{R_I \tilde R_2 \tilde R_3 
\tilde R_4}}\ ,\qquad
\gh= \frac{\sqrt{R_I ^3 \tilde R_2 \tilde R_3 \tilde R_4}}{\gii \lii
  ^3}\ ,\qquad\label{iihet1}\\
R_1&=&\gii \lii\ ,\qquad 
R_i=\frac{\gii \lii ^3}{\tilde R_j \tilde R_k}\ ,\ i,j,k=2,3,4\ .
\label{iihet2}
\end{eqnarray}
\end{subequations}
while the torus $T^2(R_5,R_6)$, common to both sides, is still at
the type II string scale. 

In the $\ell=0$ case, the dual heterotic string has a scale
$\lh=\gii$ in type II units, of the same order as the radii $R_{1,2,3,4}$;
$R_{5,6}$ on the other hand still have a type II string scale, and are
much larger (at weak type II coupling) than the heterotic scale. This
situation is therefore identical to the heterotic $n=2$ case in 
(\ref{scal2}), which did not admit a perturbative type I dual.

For $\ell\ge 1$, T-duality on the $\Kt$ manifold allows us to choose
one of the large directions as the interval of length $R_I=R\gg \lii$.
We therefore consider a regime where
\begin{equation}
R_I = R\ ,\qquad \tilde R_2 \tilde R_3 \tilde R_4 = R^{\ell-1}\ ,\qquad
R_{5}=R_6 = \lii\ ,\qquad
\gii\ll 1
\end{equation}
in units of the type IIA string length $\lii$. The parameters for the
dual heterotic string therefore scale as
\begin{equation}
\lh = \gii R^{-\ell/2}\ ,\qquad\gh=\frac{R^\frac{2+\ell}{2}}{\gii}\
,\qquad R_1=\gii\ ,\qquad
R_i = \frac{\gii}{\tilde R_j \tilde R_k}\ .
\end{equation}
A simple case by case study shows that the $\ell=1,2,4$ cases
are identical to the $n=6,4,3$ heterotic cases, up to powers of $\gii$
which we now consider of order 1. The $\ell=3$ case on the other hand
is new, since it involves, after T-duality along the direction $R_4$, 
three large directions of size $R_{2,3,\hat 4}\sim\lh (R/\lii)^{1/2}$, and 
three extra-large ones of size $R_{1,5,6}\sim \lh (R/\lii)^{3/2}$. It 
does not yield, however, any perturbative description on the type I
side. Again we see that the $n=5$ heterotic case does not appear,
since it corresponds to a strongly coupled type II theory.

We now turn to the type IIB theory,
again compactified on the model (\ref{ktt}) of $\Kt\times T^2$
for simplicity. Since T-duality on one of the circles $R_{5,6}$ 
identifies the type IIA and IIB theories, it is sufficient to
restrict our attention to the case 
where both circles are much larger
than the type II string length, but still of comparable size in order
to maintain a small gauge coupling $\gym^2=R_5/R_6$. The type IIB
theory is then equivalent to a strongly coupled heterotic theory
with parameters
\begin{subequations}
\label{hetiib}
\begin{eqnarray}
\lh &=& \frac{\giib \lii ^3}{R_6\sqrt{R_I \tilde R_2 
\tilde R_3 \tilde R_4}}\ ,\qquad
\gh= \frac{R_6\sqrt{R_I ^3 \tilde R_2 \tilde R_3 \tilde R_4}}
{\giib \lii ^3}\ ,\qquad
\label{hetiib1}\\
R_1&=&\frac{\giib}{R_6} \lii\ ,\qquad 
\hat R_6=\frac{\lii ^2}{R_6}\ ,\qquad
\tilde R_i=\frac{\giib \lii ^3}{\tilde R_j \tilde R_k}\ ,\
i,j,k=2,3,4\ ,
\label{hetiib2}
\end{eqnarray}
\end{subequations}
where now the l.h.s. refers to type IIB variables.
For $\ell=0$, the dual heterotic theory has one large dimension
of radius $R_5=R$ and five heterotic-string--size dimensions,
up to factors of $\giib\sim 1$, which corresponds to the
situation in (\ref{scal1}). For $\ell\ge 1$, we obtain again a heterotic
theory with more than two scales, and heterotic--type I duality
does not yield any valuable perturbative description.

\section{Concluding remarks}

In this paper, we studied new scenarios of TeV strings or large dimensions in
weakly coupled type II theories and related them by duality to heterotic
string compactifications with large dimensions. In particular, we described a
type IIA theory with all compactification and string scales at the TeV, but
with a tiny string coupling which explains the weakness of gravitational
interactions. We also described a type IIB theory with two large non-transverse
dimensions at the TeV and a fundamental string scale at $10^{11}$ GeV.
The main features of our discussion are summarised in Figure \ref{sum}.

\begin{figure}
\begin{center}
\begin{tabular}{|l|c|c|c|l|c|}
\hline
$n$ & $R^{-1}_{\rm H}$ & Dual & $l^{-1}_{\rm Dual}$ & Radii & QG Scale\\
\hline
1 & TeV & IIB & $10^{11}$ GeV & 2 at TeV$^{-1}$, 4 at $l_{\rm Dual}$  &
$10^{11}$ GeV\\ 
2 & TeV & IIA & TeV & all at TeV$^{-1}$ & $10^{18}$ GeV\\
3 & TeV & IIA & TeV & 2 at TeV$^{-1}$, 4 transv. at fm  & TeV\\
4 & TeV & I or IIA & TeV & 4 at TeV$^{-1}$, 2 transv. at 0.1 mm & TeV\\
5 & $>10^{6}$ GeV & I & TeV & 1 transv. at mm, 5 transv. at GeV$^{-1}$ & TeV\\
6 & $>10^{8}$ GeV & I & TeV & 6 transv. at 0.1 fm & TeV\\
6'& $>10^{8}$ GeV & IIA& $10^8$ GeV & 1 transv. at mm, 5 at $l_{\rm
  Dual}$ & $10^8$ GeV
\\ \hline
\end{tabular}
\end{center}
\caption{Weakly coupled dual descriptions of heterotic string with $n$
large dimensions or radius $R_H$. The two last columns list the size
of the internal radii in the dual theory and the scale at which 
string interactions and quantum gravity become relevant. \label{sum} }
\end{figure}

As a result, the heterotic string with $n\le 4$ large dimensions at the TeV has
a weakly coupled description in terms of type II or type I theory, as indicated
in the table. When the number of large dimensions is $n=5$ or 6, there is an
upper bound for the compactification scale because the string threshold of the
weakly coupled dual theory appears in lower energies \cite{ckm}.
The entries in the last three rows correspond to a saturation of this bound.
Moreover, the case $n=5$ is generally forbidden since the dual type I theory has
an anisotropic transverse space with one dimension very large compared to the
others; this invalidates the decoupling of the gauge theory on the brane
unless local tadpole cancellation is imposed \cite{ab}.

In particular, we showed that the first two simple type II examples above
describe the heterotic string with one or two large TeV dimensions. In fact,
these are the only two cases that have been previously considered seriously in
the context of the heterotic theory before knowing its strong coupling behavior
\cite{ia,abq}. Our analysis here showed that many of the properties and
predictions of the heterotic string for these two cases remain valid, despite
its strong ten-dimensional coupling. More precisely:
(i) the existence of KK excitations for all Standard Model gauge
bosons in $N=4$ supermultiplets, and
their absence for quarks and leptons;
(ii) the absence of visible quantum gravity effects at the TeV scale, above
which there is a genuine six-dimensional gauge theory, regulated by
the underlying type IIA or IIB theory; (iii) the possible relation
of the TeV dimensions with the mechanism of supersymmetry breaking by the
process of compactification. All soft breaking terms can then be studied
reliably in the effective field theory due to the extreme softness of the
breaking above the compactification scale \cite{soft}; (iv) 
the possibility that
the unification of low energy gauge couplings remains at the experimentally
inferred GUT scale, which is much higher than the fundamental string scale of
the weakly coupled type II theory.

Many questions and open problems remain of course to be done. Certainly, the
possibilities discussed here give new ``viable'' directions of how string
theory may be possibly connected with the description of our observed low
energy world. We note however that the strong coupling regime of the
heterotic string is traded for a strong (singular)
curvature situation in the type II framework, which is only
partially accounted for in the geometric engineering field theory approach.

\noindent {\it Acknowledgements}~: 
We would like to thank C. Bachas, S. Dimopoulos, P. Mayr and S. Shatashvili for
useful discussions.

\enlargethispage{1cm}


\begin{thebibliography}{99}

\bibitem{ia} I. Antoniadis, {\it A Possible New Dimension at a Few TeV},
Phys. Lett. {\bf B246} (1990) 377.

\bibitem{add} N. Arkani-Hamed, S. Dimopoulos and G. Dvali, 
{\it The Hierarchy Problem and New Dimensions at a Millimeter},
Phys. Lett. {\bf B429} (1998) 263, hep-ph/9803315.

\bibitem{ab} I. Antoniadis and C. Bachas, {\it Branes and the Gauge Hierarchy},
hep-th/9812093.

\bibitem{w} E. Witten, {\it Strong Coupling Expansion of Calabi-Yau Compactification},
Nucl. Phys. {\bf B471} (1996) 135, hep-th/9602070.

\bibitem{ddg} K.R. Dienes, E. Dudas and T. Gherghetta, 
{\it Extra Space-Time Dimensions and Unification},
Phys. Lett. {\bf B436} (1998) 55, hep-ph/9803466;
{\it Grand Unification at Intermediate Mass Scales Through Extra Dimensions},
hep-ph/9806292; 
Z. Kakushadze, {\it Novel Extension of MSSM and `TeV Scale' Coupling Unification},
hep-th/9811193.

\bibitem{cb} C. Bachas, 
{\it Unification with Low String Scale}, hep-ph/9807415.

\bibitem{ant} I. Antoniadis, K.S. Narain and T.R. Taylor,
{\it Higher Genus String Corrections to Gauge Couplings},
Phys. Lett. {\bf B267} (1991) 37.

\bibitem{soft} I. Antoniadis, S. Dimopoulos and G. Dvali, 
{\it Millimetre-range Forces in Superstring Theories With Weak-scale
Compactification}, Nucl. Phys. {\bf B516} (1998) 70;
I. Antoniadis, S. Dimopoulos, A. Pomarol and M. Quir{\'o}s,
{\it Soft Masses in Theories with Supersymmetry Breaking by TeV-Compactification},
hep-ph/9810410.

\bibitem{aadd} I. Antoniadis, N. Arkani-Hamed, S. Dimopoulos and G. Dvali, 
{\it New Dimensions at a Millimeter to a Fermi and Superstrings at a TeV},
Phys. Lett. {\bf B436} (1998) 263, hep-ph/9804398. 

\bibitem{st} G. Shiu and S.-H.H. Tye, 
{\it TeV Scale Superstrings and Extra Dimensions},  
Phys. Rev. {\bf D58} (1998) 106007, hep-th/9805157;
Z. Kakushadze and S.-H.H. Tye, {\it Brane World}, hep-th/9809147.

\bibitem{l} J.D. Lykken, {\it Weak Scale Superstrings}, 
Phys. Rev. {\bf D54} (1996) 3693, hep-th/9603133.

\bibitem{wi}
E.~Witten,
{\it String theory dynamics in various dimensions},
Nucl. Phys. {\bf B443} (1995) 85,
hep-th/9503124.

\bibitem{ht2} C.M. Hull and P.K. Townsend,
{\it Enhanced gauge Symmetries in Superstring Theories},
Nucl. Phys. {\bf B451} (1995) 525, hep-th/9505073.

\bibitem{grav} G.F. Giudice, R. Rattazzi and J.D. Wells,
{\it Quantum Gravity and Extra Dimensions at High-Energy Colliders},
hep-ph/9811291; E.A. Mirabelli, M. Perelstein and M.E. Peskin,
{\it Collider Signatures of New Large Space Dimensions}, hep-ph/9811337.

\bibitem{abq} I. Antoniadis and K. Benakli, {\it Limits on Extra Dimensions in
Orbifold Compactifications of Superstrings},
Phys. Lett. {\bf B326} (1994) 69;
I. Antoniadis, K. Benakli and M. Quir{\'o}s,
{\it Production of Kaluza-Klein States at Future Colliders},
Phys. Lett. {\bf B331} (1994) 313.

\bibitem{wi2} E.~Witten, {\it Some comments on string dynamics},
Proceedings of SUSY 95, hep-th/9507121.

\bibitem{vw}
C.~Vafa and E.~Witten,
{\it Dual string pairs with N=1 and N=2 supersymmetry in four-dimensions},
Nucl. Phys. Proc. Suppl. {\bf 46} (1995) 225, hep-th/9507050~;
S.~Kachru and E.~Silverstein,
{\it N=1 dual string pairs and gaugino condensation}
Nucl. Phys. {\bf B463} (1996) 369, 
hep-th/9511228.

\bibitem{mayr}
S. Katz and C. Vafa, {\it Matter from geometry},
Nucl. Phys. {\bf B497} (1997) 146, hep-th/9606086; for a recent
review, see
P.~Mayr, {\it Geometric construction of N=2 gauge theories},
Fortsch. Phys. {\bf 47} (1998) 39,
hep-th/9807096. 

\bibitem{log}
E.~Kiritsis and B.~Pioline,
{\it On $R^4$ threshold corrections in IIb string theory and (p, q) string
                  instantons},
Nucl. Phys. {\bf B508} (1997) 509, hep-th/9707018;
A.~Gregori, E. Kiritsis, C. Kounnas, N.A. Obers, 
P.M. Petropoulos and B. Pioline,
{\it $R^2$ corrections and nonperturbative dualities of N = 4 string ground
                  states},
Nucl. Phys. {\bf B510} (1997) 423, hep-th/9708062.

\bibitem{sei}
N.~Seiberg,
{\it Nontrivial fixed points of the renormalization group in six-dimensions},
Phys. Lett. {\bf B390} (1997) 169,
hep-th/9609161.

\bibitem{fpt} I. Antoniadis, N. Arkani-Hamed and S. Dimopoulos, 1998,
unpublished.

\bibitem{pw} J. Polchinski and E. Witten,
{\it Evidence for Heterotic - Type I String Duality}, 
Nucl. Phys. {\bf B460} (1996) 525, hep-th/9510169.

\bibitem{gin}
P.~Ginsparg, {\it On Toroidal Compactification Of Heterotic Superstrings},
Phys. Rev. {\bf D35} (1987) 648.

\bibitem{Aspinwall:1996mn}
P.S.~Aspinwall, {\it K3 surfaces and string duality}, hep-th/9611137.

\bibitem{Obers:1998fb}
N.A.~Obers and B.~Pioline, {\it U duality and M theory}, to appear
in Phys. Rep., hep-th/9809039; N.A.~Obers and B.~Pioline,
{\it U duality and M theory, an algebraic approach},
to appear in the Proceedings of the 2nd Conference on 
Quantum Aspects of Gauge Theories,
Supersymmetry and Unification, Corfu, 1998, hep-th/9812139.

\bibitem{ht} C.M. Hull and P.K. Townsend, {\it Unity of Superstring Dualities},
Nucl. Phys. {\bf B438} (1995) 109, hep-th/9410167.

\bibitem{to} P.K.~Townsend, {\it The eleven-dimensional 
supermembrane revisited}, Phys. Lett. {\bf B350} (1995) 184, hep-th/9501068.

\bibitem{hw} P. Ho\v{r}ava and E. Witten, 
{\it Heterotic and Type I String Dynamics from Eleven Dimensions},
Nucl. Phys. {\bf B460} (1996) 506, hep-th/9510209.

\bibitem{egkr}
S.~Elitzur, A.~Giveon, D.~Kutasov and E.~Rabinovici,
{\it Algebraic aspects of matrix theory on $T^d$},
Nucl. Phys. {\bf B509} (1997) 122,
hep-th/9707217.

\bibitem{dine}
M.~Dine and Y.~Shirman,
{\it Truly strong coupling and large radius in string theory},
Phys. Lett. {\bf B377}(1996) 36, hep-th/9601175.

\bibitem{Aspinwall:1998eh}
P.S.~Aspinwall, 
{\it M theory versus F theory pictures of the heterotic string},
Adv. Theor. Math. Phys. {\bf 1} (1998) 127,
hep-th/9707014.

\bibitem{ckm} E. Caceres, V.S. Kaplunovsky and I.M. Mandelberg,
{\it Large Volume and String Compactifications, Revisited},
Nucl. Phys. {\bf B493} (1997) 73, hep-th/9606036.

\bibitem{aq} I. Antoniadis and M. Quir{\'o}s, 
{\it Large Radii and String Unification},
Phys. Lett. {\bf B392} (1997) 61, hep-th/9609209.

\end{thebibliography}
\end{document}